\documentstyle[psfig,10pt,multicol]{mn}

\newcommand{\xtej}{\mbox{XTE~J1550$-$564}}

\title[XTE~J1550-564]{\textit{Chandra} and \textit{RXTE} Spectroscopy
of the Galactic Microquasar XTE~J1550$-$564 in Outburst}

\author[J. M. Miller et al.]{J.~M.~Miller$^1$, H.~L.~Marshall$^1$,
	R.~Wijnands$^{1,\dag}$,T.~Di~Matteo$^{2}$, D.~W.~Fox$^3$, \and
	J.~Kommers$^1$, D.~Pooley$^1$, T.~Belloni$^4$, J.~Casares$^5$,
	P. A. Charles$^6$, A.~C.~Fabian$^7$, \and M. van der Klis$^8$,
	W.~H.~G.~Lewin$^1$\\ 
	$^1$Center for Space Research and Department of Physics,
	Massachusetts Institute of Technology,\\ Cambridge, MA
	02139-4037, USA; jmm@space.mit.edu\\
	$^2$Max Planck Institute for Astrophysics, Garching,
	Karl-Schwarzschild-Str. 1, Postfach 1317, D-85741, Garching, DE\\
	$^3$Space Radiation Laboratory, California Institute of
	Technology, Pasadena, CA, 91125, USA\\
	$^4$Osservatorio di Brera, via Brera 28, Milano, I-20121, IT\\
	$^5$Instituto de Astrofisica de Canarias, 38200 La Laguna,
	Tenerife, ES\\
	$^6$University of Southampton, SO17 1BJ, Southampton\\ 
	$^7$Cambridge University Institute of Astronomy, Madingley
	Road, Cambridge CB3 OHA\\
	$^8$Astronomical Institute ``Anton Pannekoek,'' University of
	Amsterdam and Center for High Energy Astrophysics,\\ Kruislaan
	403, 1098 SJ, Amsterdam, NL\\
	$\dag$Chandra Fellow}

\begin{document}
\maketitle

\begin{abstract}
On two occasions, we obtained nearly simultaneous~ $\simeq
4$~kilosecond snapshot observations of the Galactic black hole and
microquasar XTE~J1550$-$564 with \textit{Chandra} and \textit{RXTE}
near the peak of its May, 2000 outburst.  The low-energy sensitivity
of \textit{Chandra} and the resolution of the High Energy Transmission
Grating Spectrometer (HETGS), coupled with the broad energy range and
large collecting area of \textit{RXTE}, have allowed us to place
constraints on the outburst accretion flow geometry of this source in
the ``intermediate'' X-ray state.  The 0.65--25.0~keV continuum
spectra are well-described by relatively hot ($kT \simeq 0.8$~keV)
accretion disk and hard ($\Gamma \simeq 2.3$) coronal power-law
components.  Broad, relatively strong Fe~K$\alpha$ emission line
($W_{K \alpha} \simeq 170$~eV) and smeared absorption edge components
consistent with Fe~XXV are strongly required in joint spectral fits.
The resolution of the \textit{Chandra}/HETGS reveals that the broad
Fe~K$\alpha$ emission lines seen clearly in the individual
\textit{RXTE} spectra are not due to an intrinsically narrow line.
\end{abstract}


\section{Introduction}
XTE~J1550$-$564 is a transient Galactic black hole.  At the time of
writing, four outbursts --- each separated by clear quiescent periods
--- have been observed from this source.  Radio observations during
its 1998--1999 outburst revealed optically thick emission, with
apparent superluminal motion (v$_{\rm jet} > 2c$) (Hannikainen et
al. 2000).  As a result, \xtej~was further classified as a
``microquasar.''  Recently, Orosz et al. (2002) have constrained the
mass of the black hole primary in this system through optical
observations of the source in quiescence ($9.86~M_{\odot} < M_{BH} <
11.58~M_{\odot}$).

The accretion flow geometry of Galactic black holes in outburst may
change considerably with the mass accretion rate ($\dot{m}$), which is
often assumed to be traced directly by the soft X-ray luminosity (see,
however, Homan et al. 2001).  In transient outbursts, the soft X-ray
luminosity may suddenly rise, and decay over a period ranging between
days and months; from quiescence to the outburst peak, the luminosity
may increase by a factor of $10^{6}$ or more.  In all such systems,
periods of characteristic X-ray timing and spectral behaviors and
correlated multi-wavelength properties are observed which are known as
``states'' (for reviews, see Tanaka \& Lewin 1995, Belloni 2001, Done
2002; for a critical discussion see Homan et al. 2001).  The existence
of states likely indicates that only certain accretion modes and
geometries are allowed in Galactic black hole systems.

At luminosities below $L_{X} \sim 10^{34-35}~{\rm erg}~{\rm s}^{-1}$,
the geometry is particularly uncertain; the accretion disk may be
recessed to $R_{in} \simeq 10^{2-4}~R_{g}, R_{g} = G M_{BH} / c^{2}$
(see, e.g., Esin, McClintock, \& Narayan 1997; for a competing
description see Ross, Fabian, \& Young 1999).  In contrast, near peak
outburst luminosity --- $L_{X} \sim 10^{37-38}~{\rm erg}~{\rm s}^{-1}$
--- there is general agreement that the accretion disk may extend to
the marginally stable circular orbit.  In such cases, the disk may
serve as a probe of the general relativistic regime around the black
hole.  

This regime can be explored through continuum spectroscopy (e.g.,
through the multi-colour disk blackbody model, Mitsuda et al. 1984),
studies of high frequency quasi-periodic oscillations (QPOs) in the
X-ray lightcurve (possibly associated with the Keplerian frequency at
the marginally stable circular orbit), and through studies of
Fe~K$\alpha$ line profiles and X-ray reflection spectra (Fabian et
al. 1989, Miller et al. 2002a; George \& Fabian 1991, Ross, Fabian, \&
Young 1999).  XTE~J1550$-$564 and GRO~J1655$-$40 are the only two
dynamically-constrained Galactic black holes for which each of these
tools has revealed evidence of an accretion disk at the marginally
stable orbit, and perhaps evidence of black hole spin (Sobczak et
al. 1999, Sobczak et al. 2000; Remillard et al. 1999a, Homan,
Wijnands, \& van der Klis 1999; Strohmayer 2001, Miller et al. 2001a).

We obtained two nearly-simultaneous observations of XTE~J1550$-$564
with \textit{Chandra} and \textit{RXTE} near the peak of its May, 2001
outburst, at source luminosities of $L_{X} \simeq 0.8 \times
10^{38}~{\rm erg}~{\rm s}^{-1}$ and $L_{X} \simeq 1.0 \times
10^{38}~{\rm erg}~{\rm s}^{-1}$ (0.65--10.0~keV) for $d=5.3$~kpc
(Orosz et al. 2002).  Our short $\simeq 4$~kilosecond observations
with \textit{Chandra} were designed for a much brighter source; the
sensitivity achieved does not permit detailed studies of the
Fe~K$\alpha$ line region.  However, we are able to confirm that broad
Fe~K$\alpha$ emission lines in the \textit{RXTE} spectra are not due
to an intrinsically narrow line.

\section{Outburst History and Observation}

\xtej~was discovered on 7 September 1998 (Smith 1998), by the
\textit{All Sky Monitor} (ASM) aboard the \textit{Rossi X-ray Timing
Explorer} (RXTE).  A radio (Campbell-Wilson et al. 1998) and optical
counterpart (Orosz, Bailyn, \& Jain 1998) was quickly identified.  The
ensuing outburst lasted until June 1999, and displayed some of the
most remarkable behavior yet seen in an X-ray nova, e.g.\ an initial
flare which reached a flux level equivalent to 6.8~Crab (1.5--12~keV,
\textit{RXTE}/ASM), very rapid state transitions, and strong (7\% rms)
QPOs at frequencies as high as 285~Hz (Sobczak et al. 2000, Homan et
al. 2001).

On 2 April, 2000, new source activity was noticed by the \textit{RXTE}
ASM (Smith et al. 2000).  \xtej~was active in X-rays for nearly 70
days thereafter, and reached a peak flux of $\sim$1 Crab (ASM,
1.5-12~keV); it was also seen out to 300~keV with BATSE aboard
\textit{CGRO}, and simultaneously in optical bands (Masetti \& Soria
2000, Jain \& Bailyn 2000).  Spectrally-inverted radio emission,
likely from a compact jet, is detected in the low/hard state, as well
as a possible discrete ejection event at the low/hard -- intermediate
state transition (Corbel et al. 2001).  The ASM light curve and (5--12
keV)/(3--5 keV) hardness ratio of this outburst is shown in Figure 1.
The lightcurve of the first outburst of \xtej~departed radically from
a typical profile (Homan et al. 2001), but the second outburst is more
similar to a fast-rise exponential-decay (or, ``FRED'') envelope.

Near the peak of this outburst, Miller et al. (2001a) report
simultaneous high-frequency QPOs at 188~Hz and 276~Hz.  These QPOs
appear in an approximate 2:3 ratio like the 300~Hz and 450~Hz QPOs
found in observations of GRO~J1655$-$40 (Strohmayer et al. 2001).  If
these QPOs are tied to the Keplerian frequency at the marginally
stable circular orbit, they provide evidence for black hole spin.
Tomsick et al. (2001a) and Kalemci et al. (2001) discuss the spectral
and timing properties, respectively, of the declining phase of this
outburst.

On 28 January, 2001, \textit{RXTE} found \xtej~to be in the low/hard
state, with timing noise of 40\% rms and a spectrum well-described by
a power-law with $\Gamma = 1.52$ (Tomsick et al. 2001b).  On 10
January, 2002, \textit{RXTE} again found \xtej~to be in the low/hard
state.  Belloni et al. (2002) report that the timing and spectral
properties typical of the low/hard state in this short outburst.

The second outburst of \xtej~met the trigger criteria for our approved
\textit{Chandra} AO-1 target of opportunity (TOO) program, as well as
our \textit{RXTE} AO-5 TOO program.  In all, we made seven
observations of \xtej~with \textit{Chandra} (18 observations with
\textit{RXTE}).  Of these, here we report on the first two
observations only (see Table 1), when the source was brightest.  Of
the remaining \textit{Chandra} observations, in two cases the HETGS
failed to insert, and in two other cases the source intensity was too
faint for our graded, continuous-clocking observational mode to yield
good spectra, but the source was clearly detected (for a good
discussion of continuous clocking mode, see Marshall et al. 2001).
Imaging data from a final observation will be presented in separate
work.

\section{Data Reduction}
\subsection{\textit{RXTE} Modes and Selections}
The \textit{RXTE} data we report here were obtained through pointed
observations of the proportional counter array (PCA), which consists
of five individual proportional counter units (PCUs).  Due to gain
uncertainties in PCU-0 following the loss of its propane layer, we
exclude data from this detector.  With this exception, we include data
from all layers of all active detectors (see Table 1; for observation
R1 this includes PCUs 2, 3, and 4; for observation R2 this includes
PCUs 1, 2, 3, and 4).  The data was reduced using the LHEASOFT suite
(version 5.1).  We apply the standard ``goodtime'' and detector
``deadtime'' corrections.  The background was calculated using the
``bright source'' model within LHEASOFT.  Response matrices were made
using the tool ``pcarsp.''

At the resolution of the PCA, the spectrum of the Crab nebula is known
to be a simple power-law.  We fit a spectrum from the Crab obtained on
14 May 2000 (as close to our observations as possible), with a
power-law index for the pulsar component of 1.8 (Knight, 1982) and a
column density of $3.2\times10^{21}~{\rm cm}^{-2}$ (Massaro et al,
2000).  Below 3~keV, residuals are very large, and so we fix 3~keV as
our lower fitting bound.  Similarly, above 25~keV, residuals in fits
to the Crab are large, and our data from XTE~J1550$-$564 becomes
dominated by the background flux, so we adopt this energy as an upper
fitting bound.  As we wish to address the iron line region in
XTE~J1550$-$564, we consider residuals as a function of energy in
power-law fits to this Crab observation.  We find that with the
addition of 0.4\% systematic errors in the 3--8~keV band, and 0.8\%
systematic errors in the 8--25~keV band, the reduced $\chi^{2}$
fitting statistic drops to 1.0 (similar results were obtained by
Tomsick et al. 2001a).  Therefore, we add this energy-dependent
systematic error to our PCA data from XTE~J1550$-$564.

\subsection{\textit{Chandra} Modes and Selections}
\xtej~was observed with the \textit{Chandra} HETGS using the ACIS-S
CCD array operating in continuous clocking (CC) mode.  Because
\xtej~had flared to 6.8~Crab (1.5--12~keV) during its 1998 outburst,
avoiding damage to the ACIS array was a paramount concern.  To this
end, we employed a large dither (20'' amplitude) to spread the most
intense parts of the zeroth order over many pixels.  Furthermore, the
zeroth order aim point was shifted toward the top of the chip via a
SIM-Z translation of $+$10.0mm from the nominal S3 aim point.  A
$+$1.33' Y-offset was used to place the iron K$_{\alpha}$ region of
the HEG~$-$1 order on ACIS-S3.  Finally, we blocked-out rows 367--467,
eliminating any read-out of zeroth-order photons.  This blocking acted
to limit any telemetry saturation that would result from the
zeroth-order counts.  The CC mode was employed to limit pile-up of the
dispersed spectrum and to achieve 3~msec time resolution.

As our observing mode is non-standard, the data is not well-suited to
reduction via the tools available in the CIAO suite.  We have managed
to build robust, custom software to reduce and analyse our data.  We
first correct for bad aspect times (e.g.\ slewing, see Table 1).  The
events are then examined in projected chip-x space to correct for
``hot'' pixels which report erroneously large event numbers (a list of
bad pixels is available through CIAO).  ACIS-S4 sometimes produces
``streaks'' which appear in the ACIS chip image (see
http://asc.harvard.edu); the events were removed using a 50 column
median filter, rejecting deviations greater than 3$\sigma$.  The
spacecraft dither pattern is removed by calculating the mean chip-x
position as a function of time, which is a sine wave.  We are able to
interpolate the dither-corrected chip x position of each event by
removing this sine wave.  With the dither removed, it is possible to
calculate the wavelength corresponding to each chip-x value via the
grating equation.  A linear gain correction is performed on each of
the four independent read-out and amplifier nodes on each ACIS-S chip.
Finally, we extract first-order MEG and HEG events and background
regions, and apply the appropriate response matrices.

Indeed, although we have blocked-out the zeroth-order photons, we are
able to accurately locate the position of the zeroth-order, and
therefore are able to convert accurately from chip x space to
wavelength or energy space via the grating equation.  To do this, we
make use of the fact that the ACIS chips are silicon-based, and that
silicon has an absorption edge at 1.8395~keV.  For an assumed
zeroth-order position, then, we can examine the location of this edge
in the HEG $+$1 and HEG $-$1 spectra (which are on opposite sides of
the zeroth-order position) and iterate the assumed zeroth-order
position until the edge is seen at the same energy in both (see Figure
2).

As per Marshall et al. (2001), we perform an effective area
correction, and add 5\% systematic errors to the dispersed spectra.
Due to the fact that our \textit{Chandra} observations were offset
towards the top of the ACIS-S array, less of the HEG~$-1$ and MEG~$+$1
orders were read-out in comparison to the HEG~$+1$ and MEG~$-$1
orders.  Moreover, with the spacecraft dither, the portions of the
HEG~$-1$ and MEG~$+$1 spectra dispersed far from the zeroth order were
moved onto and off of the ACIS-S array.  This is reflected in the
spectra we obtained from the truncated orders, which show significant
deviations from the shape expected based on the spectra obtained with
\textit{RXTE}.  Therefore, we only consider the HEG~$+$1 and MEG~$-$1
spectra from our \textit{Chandra} observations.

Even after rebinning by a factor of 10, flux bins below 0.65~keV in
the MEG~$-$1 spectrum are consistent with zero.  The HEG~$+$1 spectrum
suffers from low effective area below 1.0~keV, displaying a flux trend
inconsistent with the MEG~$-$1 spectrum.  These energies serve as
lower fitting bounds.  We exclude data from regions near gaps in the
ACIS-S CCD array.  In the MEG~$-$1 spectrum we exclude the
0.74-0.80~keV band.  There is another gap between 2.2--3.0~keV in this
grating order, and at higher energies the effective area of the HEG is
higher.  We therefore adopt 2~keV as an upper fitting bound for the
MEG~$-1$ order.  We exclude the 1.3--1.4~keV and 3.2--4.2~keV bands in
the HEG~$+$1 spectrum, and set 10.0~keV as an upper fitting bound
(this is the effective upper energy limit of the HEG).

\section{Analysis and Results}
The reduced spectra were analyzed using XSPEC version 11.1.0 (Arnaud
et al. 1996).  All errors quoted in this paper reflect the difference
between the best-fit value, and the value of the parameter at its 90\%
confidence limits.

We fit the spectra from the \textit{Chandra}/HETGS and the
\textit{RXTE}/PCA spectra simultaneously; spectra R1 and C1 are nearly
simultaneous, and spectra R2 and C2 are simultaneous (see Table 1).
Joint fits are desirable to characterize the continuum emission, as
the range of the \textit{Chandra}/HETGS is well-suited to measuring
the soft spectral component expected from an accretion disk, and the
\textit{RXTE}/PCA range is well-suited to contraining the hard
power-law spectral component.

Fits with a model consisting of interstellar absorption (``phabs'' in
XSPEC), a multi-colour disk (MCD) blackbody component (Mitsuda et al.\
1984), and a power-law component did not yield acceptable fits (
$\chi^{2}/\nu > 5$).  Fits with Comptonization models such as
``compTT'' (Titarchuk 1994) and the ``bulk motion Comptonization''
model (Shrader \& Titarchuk 1999) gave slightly worse fits.  This is
principally due to strong residuals in the Fe~K$_{\alpha}$ region.  We
therefore add a broad Gaussian line and smeared edge (``smedge''
within XSPEC; Ebisawa et al. 1994).  This is the same model used by
Sobczak et al. (2000) in fits to the 1998--1999 outburst of \xtej, and
to be consistent with this prior work we fix the width of the edge to
7.0~keV, and the width (FWHM) of the line to 1.2~keV.  The widths are
poorly constrained when allowed to vary (th 90\% confidence error
limits are approximately 50\% of the assumed values).  The addition of
line plus edge combination consistent with neutral iron
(E$_{line}=6.40$~keV, E$_{edge}=7.1$~keV) does improve the fit
considerably ($\chi^{2}/\nu \simeq 3$), but this is not an acceptable
fit.  When the Gaussian and smeared edge energies are allowed to
float, values consistent with helium-like Fe~XXV (E$_{line}=6.68$~keV,
E$_{edge}=8.83$~keV) are preferred statistically.  We fix these values
in our final model, and acceptable fits are achieved:
$\chi^{2}/\nu=$0.996 (R1$+$C1), $\chi^{2}/\nu=$1.017 (R2$+$C2).  As a
first test of whether or not the line is intrinsically narrow, we also
made fits with a Gaussian of zero width.  The fit obtained is not
statistically acceptable ($\chi^{2}/\nu \simeq 5$), suggesting that the
observed line is not merely instrumentally broadened.

We report the best-fit model parameters in Table 2; the data and the
best-fit model are shown in Figure 3.  We measure an equivalent
neutral hydrogen column density of
N$_{H}=8.0^{+0.4}_{-0.3}\times10^{21}~{\rm cm}^{-2}$.  Relatively high
inner disk colour temperatures are found with the MCD model
(kT$=0.790\pm0.008$~keV and kT$=0.753\pm0.008$~keV, respectively).  We
measure power-law indices of $2.36\pm0.01$ and $2.32\pm0.01$,
respectively.  The unabsorbed fluxes measured via our model are listed
in Table 3; in the energy range we consider the disk flux is
approximately 60\% of the total.  These spectral results are
consistent with the \xtej~being in the intermediate state at the time
of our observations (this is confirmed by X-ray timing results, see
Miller et al. 2001a).  The broad Fe~K$\alpha$ emission line is
relatively strong and highly significant; we measure equivalent widths
of $160\pm20$~eV and $180\pm20$~eV, respectively.  We measure the
optical depth of the smeared edge to be $1.34\pm0.07$ and
$1.39\pm0.06$, respectively.
 
For a known source distance and inclination, the MCD model gives a
measure of the inner disk radius via the model normalization.
Assuming the mean values for the black hole mass ($M_{BH} \simeq
10.7~M_{\odot}$), system inclination ($i \simeq 73.1^{\circ}$), and
distance ($d \simeq 5.3$~kpc) reported by Orosz et al. (2002), we
measure inner disk radii of $\simeq 45$~km and $\simeq 56$~km
(corresponding to $R_{in} \simeq 2.8~R_{g}$ and $R_{in} \simeq
3.5~R_{g}$, respectively).  As these values are within the marginally
stable circular orbit ($R_{in} = 6~R_{g}$) for a non-spinning
(Schwarzschild) black hole, they may imply a black hole with
significant angular momentum in XTE~J1550$-$564.  

\subsection{Testing for narrow features.}
To further examine whether or not a single emission line might merely
be smeared by the PCA response to produce the broad Fe~K$\alpha$ line
required in joint fits, we calculated the 95\% confidence upper-limits
on the strength of single-bin emission lines in the
\textit{Chandra}/HETGS spectra (see Figure 4).  Due to Auger
destruction, intermediate Fe ion species are not likely to be
observed; relatively few narrow Fe~K$\alpha$ lines (e.g., Fe~I, XXV,
XXVI at 6.40, 6.68, and 6.97~keV, respectively) are expected.  It is
clear that the upper-limits on line features at these energies are a
fraction of the $\simeq 170$~eV equivalent width of the broad lines
required in joint fits.  The spectra we obtained are of too coarse a
sensitivity to detect narrow emission or absorption features like
those found in the high/soft state of XTE~J1650$-$500 (Miller et al. 2002b).

\section{Discussion}
We have analyzed spectra from two nearly-simultaneous
\textit{Chandra}/HETGS and \textit{RXTE} observations of the Galactic
microquasar XTE~J1550$-$564.  We jointly fit the 0.65--25.0~keV
spectra with a continuum model consisting of two components: a
multi-colour disk blackbody (likely from an accretion disk) and a
power-law (likely from a corona).  This simple model described the
data better than Comptonization models.  The data strongly require the
addition of a broad Gaussian emission line and smeared edge features
consistent with Fe~XXV (see Section 3 and Table 2).  This model may be
regarded as an approximation to a full reflection model in the
0.65--25.0~keV band.

The multi-colour disk blackbody model yields a measure of the inner
accretion disk edge.  Our fits give inner disk radii of $R_{in} \simeq
2.8~R_{g}$ and $R_{in} \simeq 3.5~R_{g}$, which suggests that
XTE~J1550$-$564 may harbor a spinning black hole.  If the Shimura \&
Takahara (1995) colour correction is applied ($R_{in, ST} = f^{2}
\times R_{in, MCD}$, $f = 1.7$ for stellar-mass black holes) black
hole spin is not required.  However, this correction may be overly
simplifed.  Merloni, Fabian, \& Ross (2000) have detailed more
systematic difficulties with the inner disk estimates of this model,
but suggest that the MCD model may give acceptable inner disk measures
at high $\dot{m}$.  The 284~Hz QPO found in the 1998--1999 outburst of
this source (Homan et al. 1999, 2001) and the 276~Hz QPO found in this
outburst (Miller et al. 2001a) also suggest a spinning black hole.

Broad Fe~K$\alpha$ emission line studies may provide another means of
diagnosing the accretion flow geometry and black hole spin, as these
lines are plausibly produced by irradiation of the inner disk (Fabian
et al. 1989).  We find that lines with widths of 1.2~keV (FWHM)
adequately describe our data.  Intrinsically narrow lines (with widths
fixed at zero, as per a line with a FWHM less than the
\textit{RXTE}/PCA resolution) give significantly worse fit results.
Moreover, the \textit{Chandra}/HETGS spectra confirm that the line is
not likely due to an intrinsically narrow line which is smeared by the
resolution of the \textit{RXTE}/PCA (see Section 3 and Figure 4).  The
sensitivity achieved with our \textit{Chandra}/HETGS snapshot
observations is not sufficient to constrain the parameters of more
sophisticated line models which can estimate the inner disk extent
more directly (and thereby black hole spin, e.g. Miller et al. 2002a).
The 1.2~keV (FWHM) width we adopt in fits with a Gaussian model is not
broad enough to require black hole spin, but it does suggest
significant Doppler shifting.  Compton broadening may also contribute
to the observed line width.

Our best-fit spectral model is the same as that adopted by Sobczak et
al. (2000) in fits to \textit{RXTE} spectra from the 1998--1999
outburst.  The parameters we measure are broadly consistent with those
reported previously when XTE~J1550$-$564 was found to be in the very
high or intermediate states.  The properties of these states are quite
similar, and they are only distinguished by relative luminosity (if
multiple such states are observed, the most luminous may be called the
``very high'' state, and the others ``intermediate'' states).  The
Fe~K$\alpha$ emisson line equivalent widths we measure ($W_{K \alpha}
= 160 \pm 20~{\rm and}~180 \pm 20$~eV) are consistent with the
strongest lines reported by Sobczak et al. (2000).  We speculate that
the sensitivity and energy range of the \textit{Chandra}/HETGS spectra
have allowed for a better characterization of the continuum spectrum,
and therefore a better characterization of the line parameters as
well.

At present, few very high and intermediate states have been observed
relative to the more common high/soft and low/hard states.  The
spectra we have observed are similar to other very high and/or
intermediate states observed in the \textit{RXTE} and
\textit{BeppoSAX} era, including: XTE~J1650$-$500, GRO~J1655$-$40,
GRS~1739$-$278, XTE~J1748$-$288, and XTE~J2012$+$381 (Miller et
al. 2002, Sobczak et al. 1999, Borozdin et al. 1998, Miller et
al. 2001b, Campana et al. 2002).

\section{Acknowledgments}
We wish to acknowledge Harvey Tananbaum, Fred Seward, Jean Swank, Evan
Smith, and the \textit{Chandra} and \textit{RXTE} operations staffs,
for executing and coordinating these TOO observations.  We thank
\textit{Chandra} scientist David Huenemoerder for sharing his analysis
software expertise, and \textit{RXTE}/PCA scientist Keith Jahoda for
his generous help with the PCA detector response.  We thank Ron
Remillard for helpful discussions.  W.~H.~G.~L. gratefully
acknowledges support from NASA.  R.~W. was supported by NASA through
Chandra fellowship grant PF9-10010, operated by the Smithsonian
Astrophysical Observatory for NASA under contract NAS8-39073.  This
research has made use of the data and resources obtained through the
HEASARC on-line service, provided by the NASA--GSFC.

\clearpage

\begin{table}

\footnotesize
\begin{center}
\begin{tabular}{lllllllllll}
No. & \multicolumn{2}{l}{Obs.ID} & Date, UT & Start Time & Day (year
2000) & Exp. (s) & Good Time (s)\\ 
\hline
\hline
C1 & \multicolumn{2}{l}{680} & 05/03/00 & 21:10:40 & 122.9 & 3380 & 2670 \\ 
C2 & \multicolumn{2}{l}{681} & 05/06/00 & 12:54:09 & 125.5 & 3770 & 2170 \\ 
\hline
R1 & \multicolumn{2}{l}{50134-02-04-00} & 05/03/00 & 16:33:20 & 122.7
& 4860 & 4860\\
R2 & \multicolumn{2}{l}{50134-02-06-00} & 05/06/00 & 12:50:08 & 125.5
& 3960 & 3960\\
\hline
\hline
\end{tabular} ~\vspace*{\baselineskip}~\\ \end{center}
\caption{Observation log.  These observations were made shortly after
outburst maximum, during the ``intermediate'' X-ray state.  ``Good
Time'' is non-slewing observation time.  C1 and C2 are the
\textit{Chandra} observations, R1 and R2 are the \textit{RXTE}
observations.}
\end{table}

\begin{figure}
\begin{center}
\begin{tabular}{c}
\psfig{file=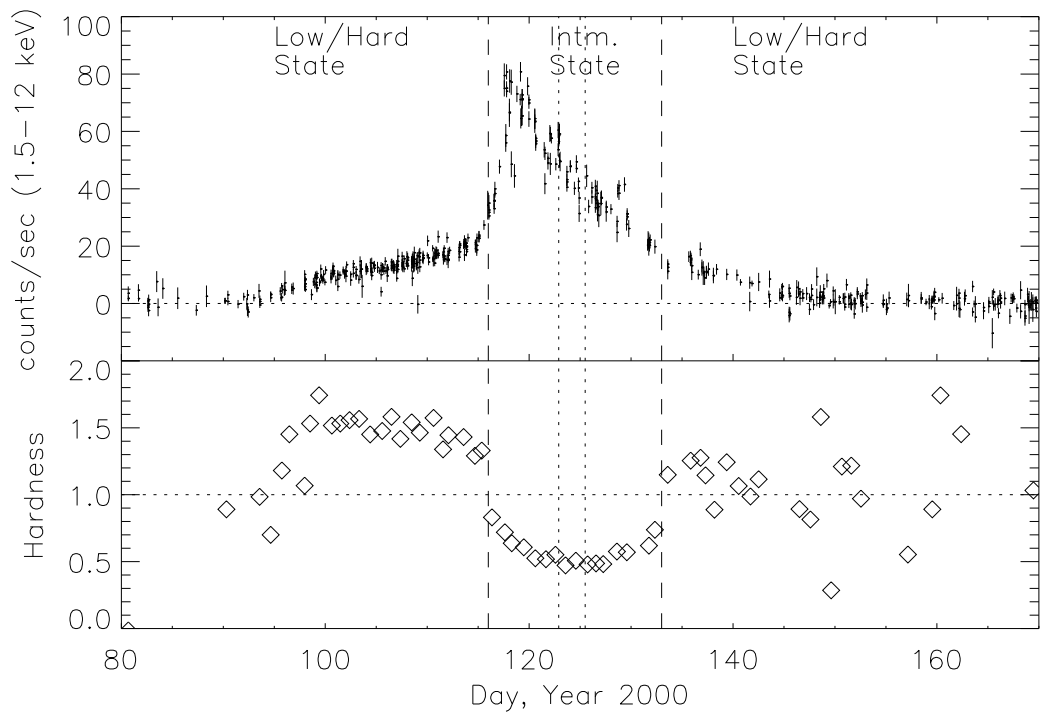,width=6.0in,height=4.0in}
\end{tabular}
\caption{The ASM 1.5-12~keV lightcurve and (5-12 keV)/(3-5 keV)
count-rate ratio for the April-May 2000 outburst of \xtej.  Dotted
vertical lines indicate the approximate start times of our
\textit{Chandra} observations.  Dashed vertical lines separate X-ray
states.  State identifications based on the ASM hardness ratio, and
radio observations reported by Corbel et al.\ (2001).}
\end{center}
\end{figure}

\clearpage

\begin{figure}
\begin{center}
\begin{tabular}{c}
\psfig{file=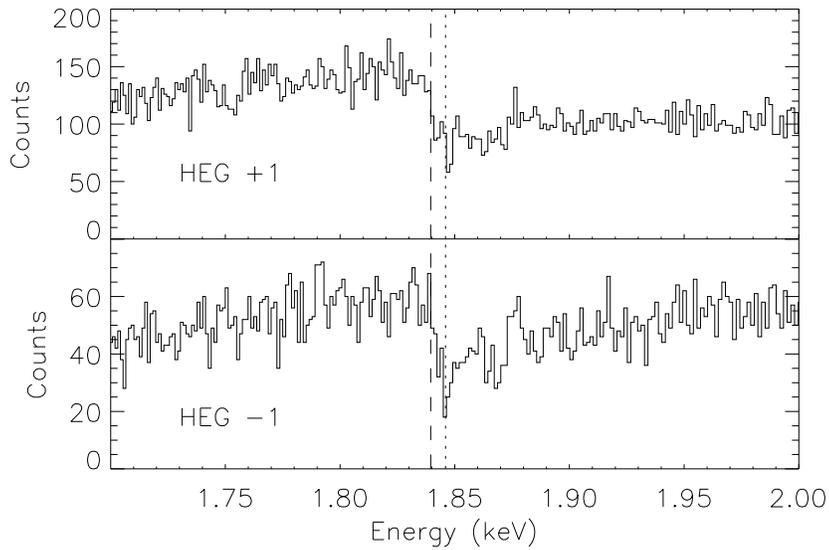,width=4.5in,height=3.0in}\\
\end{tabular}
\caption{The count spectrum of \xtej~near the instrumental silicon
absorption edge.  The dashed line is the laboratory edge energy
(1.8395~keV).  The dotted line is the edge minimum, which we can align
to within the instrumental resolution.  This provides an effective
means of establishing the zeroth-order position, and therefore energy
as a function of position on the ACIS-S array.  This is necessary as
we blocked the zeroth-order photons with a spatial window to prevent
telemetry saturation.}
\end{center}
\end{figure}

\clearpage

\begin{table}
\footnotesize
\begin{center}
\begin{tabular}{llllllllll}
Obs. & $N_{\rm H}$ &  kT$_{\rm MCD}$ &
N$_{\rm MCD}$ & $\Gamma_{\rm pl}$ & N$_{\rm pl}$ & N$_{line}$ & EW$_{line}$ & $\tau_{edge}$ & red.~$\chi^{2}$\\
 ~ & $(10^{21}~{\rm cm}^{-2})$ & (keV) & ~ & ~ & ~ & $(10^{-3})$ & (eV) ~ & ~ \\
\hline
\hline
R1$+$C1 & $8.0^{+0.2}_{-0.1}$ & $0.790^{+0.008}_{-0.008}$ & $2100^{+100}_{-100}$ & $2.36^{+0.01}_{-0.01}$ & $2.38^{+0.07}_{-0.07}$ & $5.7^{+0.7}_{-0.7}$ & $160^{+20}_{-20}$ & $1.34^{+0.06}_{-0.07}$ & 0.996 \\
\hline
R2$+$C2 & $8.0^{+0.4}_{-0.3}$ & $0.753^{+0.007}_{-0.008}$ & $3200^{+200}_{-200}$ & $2.32^{+0.01}_{-0.01}$ & $2.90^{+0.07}_{-0.08}$ & $7.7^{+0.7}_{-0.9}$ & $180^{+20}_{-20}$ & $1.39^{+0.06}_{-0.05}$ & 1.017\\
\hline
\hline
\end{tabular} ~\vspace*{\baselineskip}~\\ 
\end{center}
\caption{The results of fitting the \textit{RXTE}/PCA and
\textit{Chandra}/HETGS spectra (rebinned by a factor of 10) jointly.
There are 334 bins and 321 degrees of freedom for each joint fit.  The
model consists of photoelectric absorption, a multi-colour disk
blackbody component, a power-law component, and a broad Gaussian plus
smeared edge (``smedge'') in the iron K$_{\alpha}$ region.  The
Gaussian and smeared edge energies are fixed at 6.68~keV and 8.83~keV,
respectively, correspoinding to Fe~XXV.  Neutral Fe features and
spectral models without a broad line and edge give significantly worse
fits to the data.}
\end{table}

\begin{table}
\footnotesize
\begin{center}
\begin{tabular}{llllll}
Obs. & Range & Total & MCD & Power-law & Line \\
 ~ & (keV) & ($10^{-8}$~cgs) & ($10^{-8}$~cgs) & ($10^{-8}$~cgs) & ($10^{-11}$~cgs)\\
\hline
\hline
R1$+$C1 & 0.5--10.0 & $2.4^{+0.2}_{-0.1}$ & $1.5^{+0.1}_{-0.1}$ & $0.90^{+0.02}_{-0.03}$ & $6.1^{+0.7}_{-0.7}$ \\
R1$+$C1 & 0.65--25.0 & $2.3^{+0.1}_{-0.1}$ & $1.4^{+0.1}_{-0.1}$ & $0.89^{+0.02}_{-0.03}$ & $6.1^{+0.7}_{-0.7}$ \\
\hline
R2$+$C2 & 0.5--10.0 & $3.0^{+0.2}_{-0.2}$ & $1.9^{+0.1}_{-0.1}$ & $1.12^{+0.02}_{-0.03}$ & $8.3^{+0.7}_{-0.8}$ \\
R2$+$C2 & 0.65--25.0 & $2.9^{+0.1}_{-0.2}$ & $1.7^{+0.1}_{-0.1}$ & $1.12^{+0.03}_{-0.04}$ & $8.3^{+0.7}_{-0.8}$ \\
\hline
\hline
\end{tabular} ~\vspace*{\baselineskip}~\\ 
\end{center}
\caption{The unabsorbed fluxes measured via the fits detailed in Table
2.  We quote fluxes for both the ``standard'' soft X-ray band
(0.5--10.0~keV) and the full fitting range (0.65--25.0~keV) for joint
fits to the \textit{Chandra} and \textit{RXTE} data; they are very
similar.  For a distance of 5.3~kpc (Orosz et al. 2002), \xtej~was
observed at luminosities of approximately $8\times10^{37}$ and
$1\times10^{38}$ ergs/sec (0.65--10~keV) in the first and second
observations, respectively.}
\end{table}

\clearpage

\begin{figure}
\begin{center}
\begin{tabular}{c}
\psfig{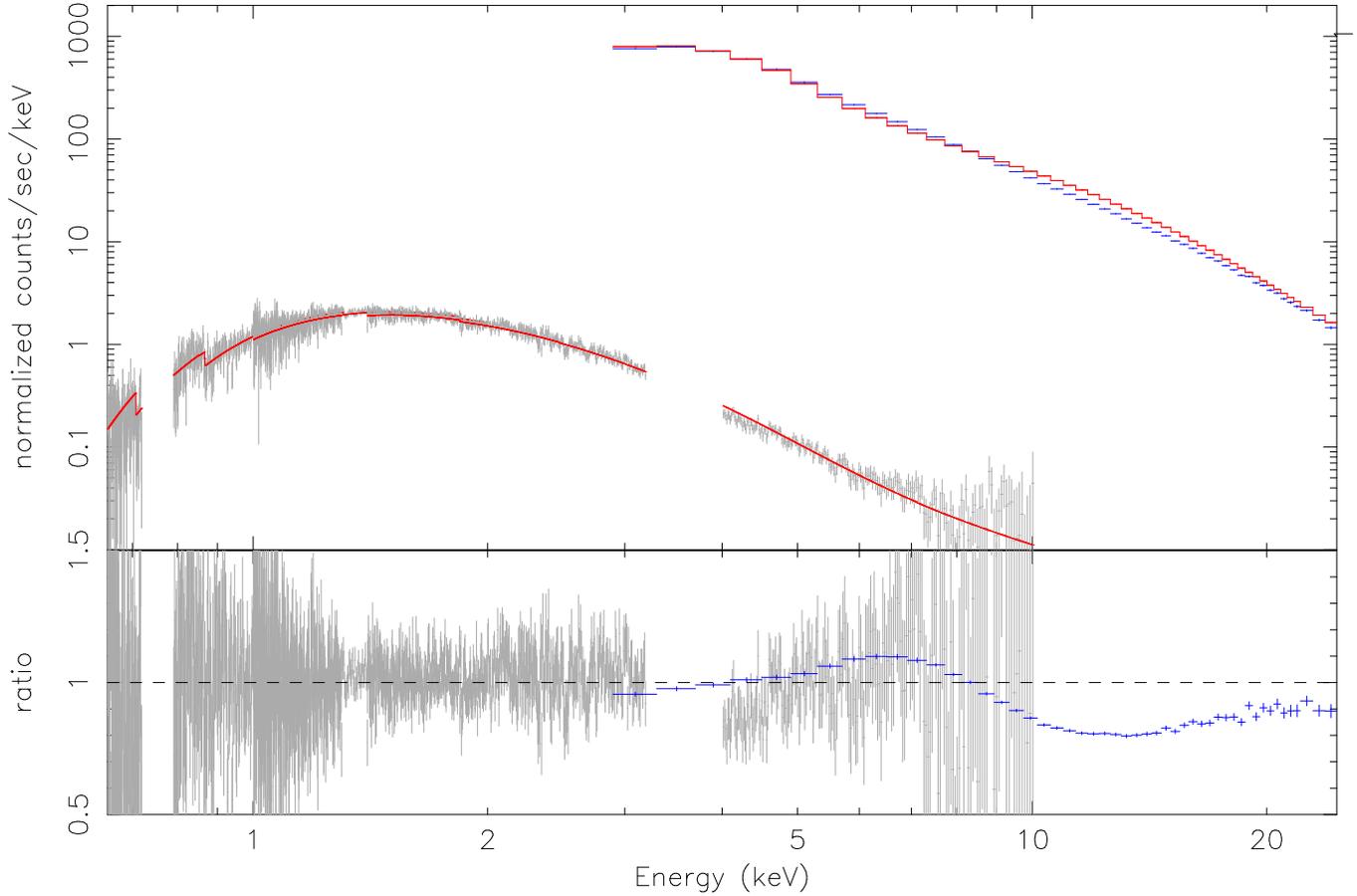}
\end{tabular}
\caption{The \textit{Chandra} and \textit{RXTE} spectra (C1 and R1)
fit with the model described in Table 2, and the resulting data/model
ratio.  The \textit{Chandra} data are shown in gray and cover the
0.65--10.0~keV band, and the \textit{RXTE} data in blue and cover the
3.0--25.0~keV band; the model is shown in red.  The Gaussian line
normalization and smeared edge depth have been set to zero to show
their strengths more clearly.  For visual clarity, the
\textit{Chandra} spectra have been rebinned by a factor of two and are
not plotted with systematic errors.  Gaps in the \textit{Chandra}
spectra are due to gaps in the CCD array which could not be recovered
in this observing mode.  The spectra and data/model ratio obtained
from fits to the observation made 2.5 days later (C2 and R2, see Table
1 and Table 2) are indistinguishable in this representation.}
\end{center}
\end{figure}

\clearpage

\begin{figure}[b]
\begin{center}
\begin{tabular}{c}
\psfig{file=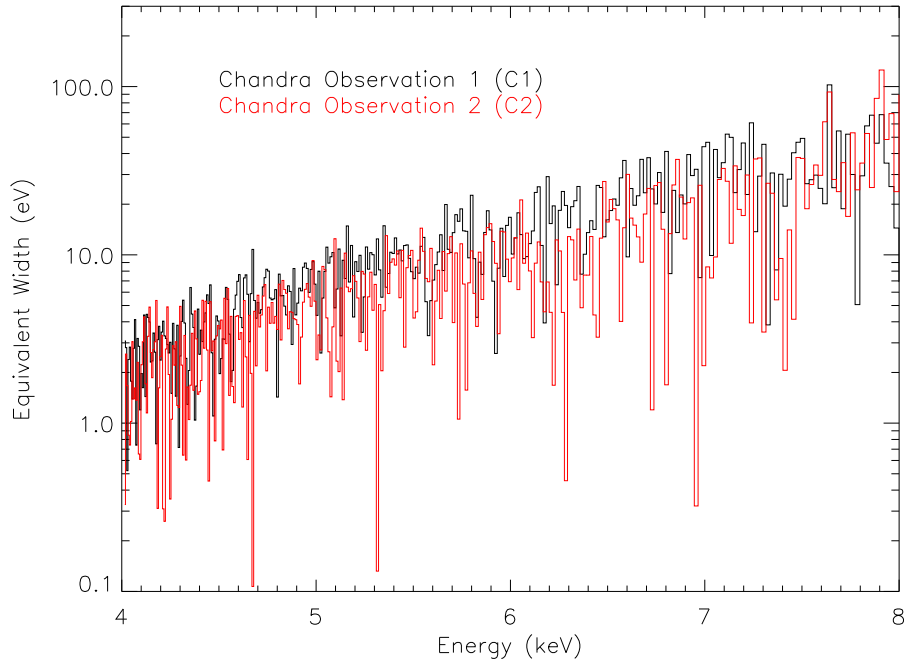,width=5.0in}
\end{tabular}
\caption{The 95\% confidence upper-limits on the equivalent width of
single-bin emission lines in the Fe~K$\alpha$ line region.  In
calculating the upper-limits, broad emission line and smeared edge
components are not included in the fit.  It is clear that the $W_{K
\alpha} \simeq 170$~eV line required in joint fits is not likely due
to an intrinsically narrow line, or even a few lines.}
\end{center}
\end{figure}

\end{document}